\documentclass[a4paper,11pt]{article}
\pdfoutput=1 
\usepackage{jcappub}
\usepackage[T1]{fontenc}
\usepackage{cleveref}
\usepackage{caption}
\usepackage{subcaption}
\newcommand{\vect}[1]{\boldsymbol{#1}}

\title{\boldmath No evidence for missing covariance in the Pantheon+ SuperNova sample distance moduli}

\author[a,b]{Bohdan Bidenko,}
\author[a]{L\'eon V. E. Koopmans,}
\author[b]{P. Daniel Meerburg}

\affiliation[a]{Kapteyn Astronomical Institute, University of Groningen, PO Box 800, NL-9700 AV Groningen, the Netherlands}
\affiliation[b]{Van Swinderen Institute for Particle Physics and Gravity, University of Groningen,
Nijenborgh 4, 9747 AG Groningen, The Netherlands}

\emailAdd{b.bidenko@astro.rug.nl}

\abstract{
Inspired by the discussion in the community on possible hidden systematic errors in late universe cosmological probes and non-trivial physical models developed to reduce the Hubble tension, we investigate the Pantheon and Pantheon+ SNe samples for possible deviations from the original $\Lambda$CDM analysis. To simultaneously account for possible systematics or deviations from $\Lambda$CDM, we adopt Gaussian processes to model additional covariance while making no further assumptions on their origin. We explore both stationary and non-stationarity corrections to the covariance. 
While small changes in the inferred cosmological parameters $H_0$ and $\Omega_{m}$ can occur, we find no statistically significant evidence for missing covariance. We find an upper limit for the Gaussian processes amplitude $\sigma < 0.031$ mag with $95\%$ confidence, which corresponds to $20\%$ of the average statistical error in the Pantheon+ sample.
The strongest effect we find on the inferred cosmological parameter posterior can reduce the statistical significance of the Hubble tension between Pantheon+ and Planck estimates from 5.3$\sigma$ to 4.5$\sigma$. 
Therefore, we conclude that the SN cosmological parameter inference is robust against the analysis modifications studied in this work.
}

\begin{document}
\maketitle
\flushbottom
\section{Introduction}
\label{sec:intro}
Significant improvements in the progression of modern-day cosmological probes allow us to measure the expansion of the Universe rate with sub-percent precision.
However, not all cosmological probes converge to the same value.
The discordance between probes based on the processes occurring at the early stages of the universe's evolution and probes using late universe astrophysical processes has reached a high level of statistical significance that cannot be ignored.
The Planck collaboration baseline analysis results in a constraint of $H_0 = 67.36 \pm 0.54 \mathrm{\ km\ s}^{-1} \mathrm{\ Mpc}^{-1}$ \cite{ref:plk:param}. 
This measurement is in agreement with other observations of the Cosmic Microwave Background (CMB), such as space-based observations by WMAP, as well as with recent ground-based observations by ACT \cite{ref:wmap,ref:actdr4}. 
Values determined by the CMB are supported by constraints obtained with measurement of the early-universe baryonic acoustic oscillation imprint on the large-scale structure of the universe. 
The BOSS and eBOSS surveys, in combination with constraints on the baryon matter density determined from the primordial deuterium abundance, suggest $H_0 = 67.33 \pm 0.98 \mathrm{\ km\ s}^{-1} \mathrm{\ Mpc}^{-1}$ \cite{ref:bosseboss}.

These values are in strong statistical disagreement with measurements based on the estimation of luminosity distance to type Ia supernovae (SNe). The inferred value of $H_0 = 73.04 \pm 1.04 \mathrm{\ km\ s}^{-1} \mathrm{\ Mpc}^{-1}$ was estimated by the SH0ES collaboration using the revised Pantheon+ SN sample based on the three-rung calibration distance ladder procedure \cite{ref:sh0es2022,ref:pan2022}. This implies that the statistical significance of this tension is at $4.8\sigma$ for the Planck collaboration value and $4\sigma$ for BOSS+eBOSS values.
Including higher redshift SNe from the complete Pantheon+ sample result in an estimate of $H_0 = 73.53 \pm 1.05 \mathrm{\ km\ s}^{-1} \mathrm{\ Mpc}^{-1}$ and the corresponding tension increase to $5.2\sigma$ and $4.3\sigma$, respectively, for Planck and BAO inferred values.
At the same time, this value is in agreement with other late universe cosmological probes. For example, distance measurements to megamaser-hosting galaxies with an inferred value of $H_0 = 73.9 \pm 3.0 \mathrm{\ km\ s}^{-1} \mathrm{\ Mpc}^{-1}$ that correspond to a $2 \sigma$ tension with the Planck probe \cite{2020ApJ...891L...1P}. The estimate based on infrared surface brightness fluctuation measurements is $H_0 = 73.3 \pm 2.5 \mathrm{\ km\ s}^{-1} \mathrm{\ Mpc}^{-1}$ with a corresponding $2.3\sigma$ tension \cite{2021ApJ...911...65B}, and strong gravitational lensing based measurements yield $H_0 = 73.3 \pm 1.8 \mathrm{\ km\ s}^{-1} \mathrm{\ Mpc}^{-1}$ with a corresponding $3.1\sigma$ tension \cite{2020MNRAS.498.1420W}.  For a detailed review, see e.g. Ref. \cite{2021APh...13102605D}.

Various cosmological probes, both early and late universe, have undergone extensive testing to explore potential extensions of their baseline analysis, which can impact the constraints on cosmological parameters in diverse ways.
Proposed solutions for the Hubble tension through modifications in the early universe include, for example, a varying effective electron mass with the possible addition of spatial curvature \cite{ref:eumod1} or an additional dark energy component acting on the early stages of the universe's expansion \cite{ref:eumod2}. An extensive review of various model modifications can be found in Ref. \cite{ref:h0olympics} and Ref. \cite{ref:divalh0solutions}.
Distance ladder based estimates have been criticized for their potential sensitivity to different treatments of the analysis of Cepheids used for the calibration and assumptions on the reddening law used for SN brightness calibration \cite{ref:red}.
The usage of different calibrators such as the Tip of the Red Giant Branch or Mira variable stars, as well as other differences in the analysis pipelines, can lower the tension \cite{ref:trgb,ref:mira,2020arXiv200710716E}.
The Pantheon sample of SNe also seems to feature redshift evolution of cosmological parameters. This apparent evolution shows up when the upper limit on the used SNe is varied \cite{ref:zmax_vs_omegam}, and with the use of various redshift bins \cite{ref:zev}. The statistical significance of these effects was shown to be at the $2\sigma$ level and is reduced in the revised Pantheon+ sample \cite{ref:pan2022}.

Motivated by these hints of trends in the data, we search for evidence of possible unaccounted systematic errors, or deviations from the standard cosmological model in the SNe datasets with minimal assumptions about the nature of underlying processes. For this reason, we model any deviation by a Gaussian process (GP) which is applied as an additive component to the covariance matrix. GPs excel in modeling physical processes with unknown parametrization due to their non-parametric nature, flexibility, and ability to capture complex relationships without assuming a specific functional form. \cite{ref:rasmussen2006gaussian}. Our implementation of the additional covariance is sensitive to the deviations of expansion history from the standard model prediction, as well. 

The structure of the paper is as follows. In \cref{sec:data}, we describe Pantheon and Pantheon+ SNe datasets used in this work. In \cref{sec:method}, we describe the methodology of cosmological parameter inference with these SNe samples and our implementation of the search for deviations from the baseline analysis. In \cref{sec:results}, we present constraints on the additional covariance obtained in this work and discuss the impact on the cosmological parameter inference and conclude in \cref{sec:concl}.
\section{Data \label{sec:data}}
\begin{figure}
     \centering
     \begin{subfigure}{0.495\textwidth}
         \centering
         \includegraphics[width=\textwidth]{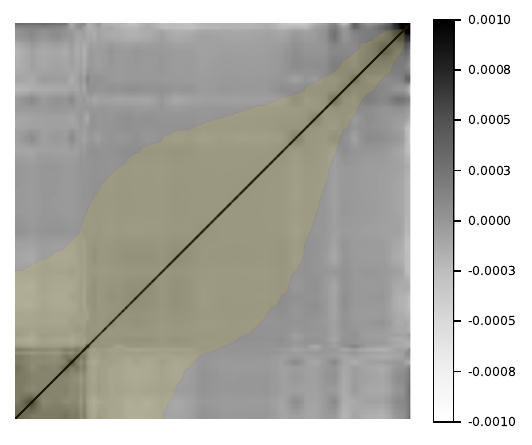}
     \end{subfigure}
     \hfill
     \begin{subfigure}{0.495\textwidth}
         \centering
         \includegraphics[width=\textwidth]{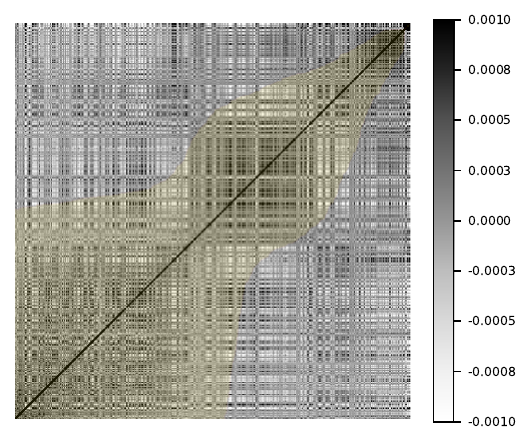}
     \end{subfigure}
        \caption{Comparison of Pantheon and Pantheon+ covariances. Covariance matrices are presented for redshift sorted dataset, with bottom-left corresponding to low redshift SNe and top-right to high redshift SNe. Covariance of SNe separated by $<0.18$ in the redshift space is highlighted with yellow color (see \cref{sec:results}).}
        \label{fig:covmats}
\end{figure}

In this work, we will search for the presence of additional covariance in the distance moduli of the Pantheon \cite{ref:pan2018} SN sample and its updated version Pantheon+ \cite{ref:pan2022}. In this section, we briefly describe these two SN samples.
\subsection{Pantheon}
The Pantheon sample contains measured lightcurves of 1048 SNe in the redshift range $0.01<z<2.3$ \cite{ref:pan2018}. This sample consists of data compiled from 12 surveys. 
Similar to the approach used by the SH0ES collaboration in \cite{ref:sh0es2016} and the cosmological parameter inference in the original 
Pantheon analysis \cite{ref:pan2018}, we apply a lower redshift cut of $z>0.023$ to minimize the impact of possible unaccounted peculiar motions of SN host galaxies.
This reduces the sample to a total of 988 SNe with one light curve measurement for each SN. 
The Pantheon sample does not provide calibration of the absolute brightness of SNe. Therefore, we adopt the results obtained with the three-rank distance ladder calibrations presented in Table 5 of \cite{ref:sh0es2016}.
This calibration is based on the calibration of SNe with Cepheids in 19 host galaxies, while Cepheids are calibrated in three nearby galaxies based on well-established distance measurements and using parallaxes of Milky Way Cepheids. 
Therefore, in relation to the original Pantheon sample analysis, we treat the calibration dataset of SNe as independent from the Hubble flow supernovae in this work and use the aforementioned constraints on SN brightness as independent from the Hubble flow SN measurements. This calibration is necessary to obtain $H_0$ constraints but does not have an appreciable impact on $\Omega_m$ constraints in $\Lambda$CDM cosmology.

\subsection{Pantheon+}
The Pantheon+ sample is the successor to the original SN sample, which was analyzed in a joint effort with the SH0ES team.
The Pantheon+ sample contains 1701 measurements of 1534 SNe in the range of $0.001<z<2.26$. 
Similar to Pantheon+ and SH0ES 2022, we apply a lower redshift cut $z>0.01$ which limits the sample to 1590 measurements of 1473 SNe \cite{ref:pan2022,ref:sh0es2022}. 
This redshift cut is motivated by the strong impact of peculiar velocities in the low-redshift range and the impact of a volumetric bias.
The absolute brightness of SNe in the Pantheon+ sample is calibrated based on a similar three-rung distance ladder approach. The improved calibration relies on 42 SNe and is further improved by using a higher number of Cepheids observed in calibration-anchor galaxies \cite{ref:sh0es2022}.
The calibration of the SNe brightnesses in the Pantheon+ sample was performed in collaboration with the SH0ES team. Therefore, in the part of our work dedicated to the study of the Pantheon+ sample, we will use the full covariance matrix between the calibration set and the Hubble flow set.

In addition, an important improvement in the updated analysis is the estimation of systematic errors. For the Pantheon sample, systematic errors related to the covariance are computed using binned data, while for the Pantheon+ analysis an estimation of the covariance for individual SN observations is implemented. This considerably improves the estimation of the error budget and leads to tighter constraints on cosmological parameters \cite{2021ApJ...912L..26B}. The two covariance matrices for the redshift-sorted samples are shown in \Cref{fig:covmats}.
In this work, we modify these covariances with an additional GP, in search of hidden systematic errors and/or deviations of cosmological model from the standard $\Lambda$CDM model.
\section{Methodology \label{sec:method}}

In this section, we describe our analysis pipeline step by step and the assumptions that we make. 
First, we provide a summary of the standard approach to the use of SNe as a cosmological probe. 
Second, we describe how we introduce modifications to the covariance matrices.
Last, we describe how we perform a model comparison, and describe the parameter ranges of the priors and the codes used for the Bayesian parameter inference.    

\subsection{Baseline cosmological parameter inference \label{sec:baseline}}
Luminosity distance measurements with standard candles is one of the most powerful probes of the late-time expansion of the Universe. 
In our inference of the cosmological parameters, we use the standardized observed SN brightness values $m_b$ corrected for the intrinsic scatter provided by the Pantheon collaboration.
The observed SN brightness $m_b$ is related to the absolute brightness $M_b$ and the luminosity distance, as follows:
\begin{equation}
 {m_b} = {\mu}({z}) + {M_b}
\,,
\qquad
{\mu}({z}) = 5 \log ({D_L}({z})) + 25,
\label{eq:distmod}
\end{equation}
where the luminosity distance $D_L$ measured in Mpc, depends on the cosmological model and the redshift of observed SN.
For a flat homogeneous and isotropic universe, it is defined only by the evolution of the universe expansion rate:
\begin{equation}
D_L(z) = c (1+z) \int^{z}_{0} \frac{a(z') d z'}{\dot{a}(z')} = c (1+z) \int^{z}_{0} \frac{d z'}{H(z')}.
\label{eq:dist}
\end{equation}
Here $a(z)$ is the scale factor and $H(z)$ is the Hubble parameter.
In this work, we consider only constraints on the flat $\Lambda$CDM cosmological model at low redshifts. Therefore, the late universe distances can be described using only the Hubble constant $H_{0}$ and the total matter energy density parameter $\Omega_{M}$: 
\begin{equation}
H(z) = H_0 \sqrt{\Omega_{M} (1+z)^{3} + (1 - \Omega_{M})}.
\label{eq:hubble}
\end{equation}
 
For cosmological parameter inference, Pantheon and Pantheon+ analyses are using observed SN brightnesses corrected for multiple factors, such as the color and stretch parameters of observed SN lightcurves, properties of host galaxies and selection biases. For details of these corrections and estimation of the corresponding systematic errors, we refer to the original Pantheon and Pantheon+ papers \cite{ref:pan2018,ref:pan2022}.
We follow the same cosmological analysis approach, using the corrected observed brightnesses as data and the corrected absolute brightness parameter $M_b$ as the only nuisance parameter. 

In this work, we are not concerned with any specific possible calibration issues related to the distance ladder calibration and SN brightness corrections, but rather look for their possible imprint on the corrected brightnesses of the SNe sample.

\subsection{Additional covariance}
We test the presence of possible deviations from the baseline analysis.
In context of many discussions about sources of possible unknown processes affecting SN absolute brightness, possible hidden systematic errors or biases in the analysis pipeline, and cosmological processes altering the scale evolution of the universe, we want to test the presence of these effects with minimal assumptions on their (physical) origin but rather focus on their statistical properties.
Therefore, we choose to capture any unmodeled effects via a GP, which introduces additional covariance but equivalently also yields a maximum-a-posteriori (and marginalised) functional form for the deviation of the data from the standard form as given by \Cref{eq:distmod,eq:dist,eq:hubble}. 
We assume the  GP to have mean zero and be stationary. We chose to use the Mat\'ern class of covariance functions to describe GP due to its simple but flexible parametrization with just three parameters. These processes are described by the following function:
\begin{equation}
    \label{eq:MatKer}
    C_{\nu}(\Delta z) = \sigma^{2} \frac{2^{1-\nu}}{\Gamma(\nu)}\left( \sqrt{2\nu}\frac{\Delta z}{d} \right)^{\nu} K_{\nu}\left(\sqrt{2\nu}\frac{\Delta z}{d}\right), 
\end{equation}
where $\Gamma$ is the gamma function and $K_{\nu}$ is the modified Bessel function of the second kind, $d$ is the characteristic scale of the covariance, $\nu$ defines shape of the covariance function, and $\Delta z$ is the distance between two supernovae in redshift space \cite{ref:mat,ref:mat2}. In appendix \ref{app:nonstat}, we discuss possible non-stationarity of the GP, while in appendix \ref{app:altk} we discuss the use of alternative kernels.

In our analysis framework, we define the new estimated covariance matrix $C_{\mathrm{tot}}$ as the sum of the covariance matrix provided by the Pantheon collaboration $C_{\mathrm{Pantheon}}$ and missing covariance captured by GP as $C_{\nu}$. Therefore, we assume it to be independent: 
\begin{equation}
    C_{\rm tot} = C_{\rm Pantheon} + C_{\nu}(\Delta z,\sigma,\nu,d).
\end{equation}
This implementation has some limitations. For example, the Mat\'ern kernel covariance is positive defined, therefore it cannot take into account negative covariance and, hence, anti-correlations.
A detection significance of such a process might be lowered if the original Pantheon covariance is overestimated, which can happen if the modeled SNe brightness scatter processes mimic residuals caused by the underlying GP.

Our modified model contains two $\Lambda$CDM parameters, the SN absolute brightness, three parameters that describe the additional GP covariance and the prior distributions of these parameters.
These six parameters are estimated simultaneously using a standard Bayesian inference with the following log-likelihood: 
\begin{equation}
    \ln\mathcal{(L)} = -\frac{1}{2} \vect{\delta}(H_{0},\Omega_{m},M,\vect{z})^T C^{-1}_{\rm tot}(\sigma,d,\nu) \vect{\delta}(H_{0},\Omega_{m},M,\vect{z}) - \frac{1}{2}\ln(\det C_{\rm tot}(\sigma,d,\nu)).
\end{equation}
Here, $\vect{\delta}$ is the vector with the residuals defined by the difference between the predicted baseline model observed SN brightnesses $m_b$ and the observed one, and $C_{\rm tot}$ is the modified covariance matrix.
The second normalisation term is usually not used in cosmological parameter inference, because covariance matrices are typically treated as constant.
In our modified model, however, this term is necessary to constrain the additional covariance amplitude parameter.
For a correct model comparison, we also include this term in the baseline analysis. 

\subsection{Priors and model comparison}
We will determine the preferred model based on the ratio of the Bayesian evidence estimated for the two models, known as the Bayes factor, with the evidence defined as: 

\begin{equation}
    \mathcal{Z} = \int_{-\infty}^{\infty} \mathcal{L}( \mathrm{data}\mid\theta) p (\theta ) d \theta.
\end{equation}
Here $p(\theta)$ is the prior on the model parameters and the integral is taken over the full parameter space $\theta$. In this formalism, positive $\Delta \log \mathcal{Z} = \log \mathcal{Z}_1 - \log \mathcal{Z}_2$ would mean preference of the first model with the probability that can be interpreted using Jeffrey's scale \cite{ref:ev2}.
\begin{table}[t]
\centering
\begin{tabular}{ | c | c | c | } 
  \hline
  parameter &  lower limit  &upper limit \\ 
  \hline
  $H_{0},\mathrm{\ km}/\mathrm{s}/ \mathrm{Mpc}$ & 65 & 85 \\ 
  $\Omega_{M}$ & 0 & 1 \\ 
  $M$ & -19.6 & -18.9 \\ 
  \hline
  $\sigma$, mag & $10^{-5}$ & 0.08 \\ 
  $d$ & $10^{-5}$ & 1 \\ 
  $\nu$ & 0.5 & 4.5 \\ 
  \hline
\end{tabular}
  \caption{\label{tab:prior} Model parameter prior ranges used in this work. The first block represent baseline analysis parameters and the second block shows parameters of the additional Gaussian process. All priors have uniform probability distribution. }
\end{table}

We adopt top-hat flat priors for all parameters in our analysis. The same set of priors is applied to both the baseline analysis and the analysis incorporating an additional GP for both datasets, as summarized in \Cref{tab:prior}. The shared prior ranges are: $65$--$85 \mathrm{\ km\ s}^{-1} \mathrm{\ Mpc}^{-1}$ for the $H_0$, $0$--$1$ for $\Omega_m$ according to the flat $\Lambda$CDM model limits, and $-19.6$ to $-18.9$ mag for $M$, which exceeds the currently known constraints significantly. For the GP parameters, we set the amplitude parameter $\sigma$ ranging from $10^{-5}$ to $0.08$ mag, where the lower limit is imposed to avoid a coordinate singularity at zero amplitude, while the upper limit is motivated by the scale of the diagonal components of the Pantheon covariance. The scale parameter $d$ is assigned priors from $10^{-5}$ to $1$, with the lower limit determined by the redshift separation between Supernovae (SNe) in the sample, and the upper limit encompassing a range that includes over $98\%$ of the measured SNe in the datasets. Finally, the prior range for the $\nu$ parameter is set from $0.5$ to $4.5$, with the lower limit chosen to avoid regions of the parameter space with excessively steep kernel realizations, and the upper limit selected to mitigate volumetric effects and exclude parameter regimes where the smoothness of the kernel remains nearly unchanged.

To obtain a posterior distributions, we use the MCMC sampler from CosmoMC \cite{ref:MCMC1,ref:MCMC2} implemented in the Cobaya package \cite{ref:Cobaya1,ref:Cobaya2} and the PolyChord nested sampler for the Bayesian evidence estimation \cite{ref:PC1,ref:PC2}.
Our likelihood implementation is based on the publicly available CosmoSIS Pantheon+ likelihood. Our version of the likelihood code was tested for the baseline analysis and is in full agreement with the published Pantheon+ collaboration MCMC chains.
All modified likelihoods and resulting MCMC chains presented in this work are publicly available\footnotemark[1]{}\footnotetext[1]{\href{https://github.com/bidenkobd/Pantheon_with_GP}{https://github.com/bidenkobd/Pantheon\_with\_GP}}. The posterior distributions were obtained and analysed with the GETDIST python package \cite{ref:getdist}.

\section{Results \label{sec:results}}
In the first part of this section, we examine the SNe observed brightness residuals from the Pantheon and Pantheon+ datasets under the standard $\Lambda$CDM model inference, specifically focusing on their correlation features. Subsequently, we present the constraints derived with the additional covariance and analyze it's influence on the inference of cosmological parameters. Finally, we calculate the Bayesian evidence ratio and draw conclusions regarding the preferred model.

\subsection{Baseline analysis}
\begin{figure}[t]
\centering 
\includegraphics[width=6in]{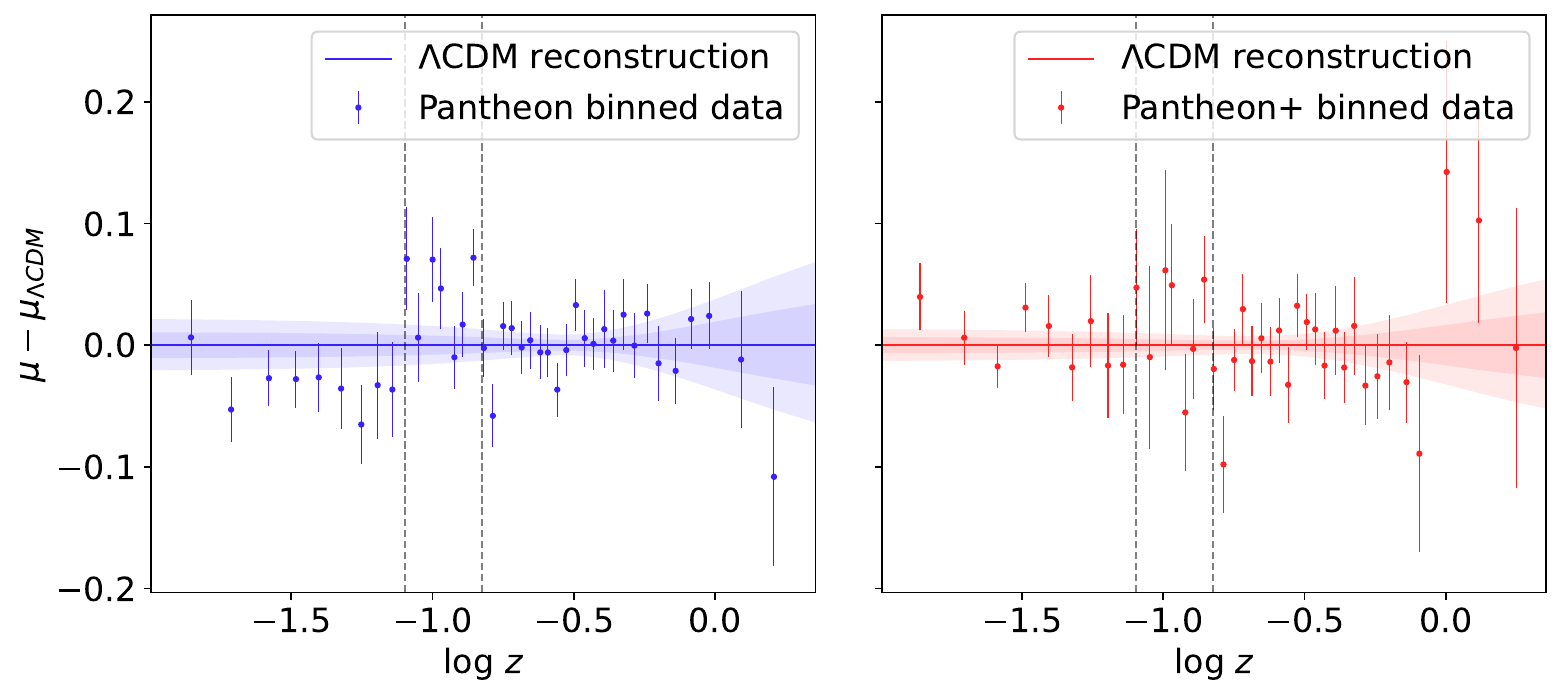}
\hfill
\caption{\label{fig:res}The left panel shows the binned residuals of  the distance moduli measured in the Pantheon SN sample with respect to the reconstructed $\Lambda$CDM cosmology distances using the SH0ES absolute brightness calibration. Dark and bright areas represent the 68\% and 95\% confidence levels of the reconstructed distances, respectively. The right panel represents the same for the Pantheon+SH0ES data. Vertical dashed gray lines are placed at $z=0.08$ and $z=0.15$. They separate regions of high correlation in the Pantheon sample residuals.}
\end{figure}

Constraints on the $\Lambda$CDM model cosmological parameter from Pantheon and its combination with SH0ES SNe calibration and the Pantheon+ sample were reported in the original papers and extensively studied in the literature \cite{ref:pan2018,ref:pan2022,ref:zev,ref:zmax_vs_omegam}.  We focus in the baseline analysis on the behavior of data residuals. In \Cref{fig:res}, we show residuals of the measured SN observed brightness with respect to the posterior prediction in a $\Lambda$CDM parameter constraints.
Blue points on the left represent redshift-binned data of the Pantheon sample which is provided as one of the original data products. The red points on the right apply a similar binning to the Pantheon+ data. 
It is worth noting that the zero level of these two plots correspond to the mean posterior prediction of Pantheon (left) and Panthoen+ (right), which are different due to slight differences in the inferred cosmological parameter constraints.

One of the motivations for the search for the additional covariance can be seen in the left panel of \Cref{fig:res}. In the  redshift range $z<0.08$ (left from the first dashed vertical line) most of the binned measured SNe brightnesses are consistently below the mean value of inferred $\Lambda$CDM model. 
At the same time, in the redshift range $0.08<z<0.15$ (between dashed lines) the residuals tends to be positive. 
For the higher redshift SN distance residuals there are no obvious apparent correlations and the data appears consistent with the model.
This kind of distribution has led to the inferred redshift evolution of cosmological parameters on a level of $2\sigma$ statistical significance \cite{ref:zmax_vs_omegam,ref:zev}.
However, the presence of the correlated residuals is not unexpected, taking into account that the error bars in \Cref{fig:res} represent only diagonal components of the covariance matrix. This can be seen as block structure of off-diagonal component seen in the left panel of \Cref{fig:covmats}.  
This has led us to explore the hypothesis that some part of the covariance is missing.
This additional covariance could represent systematic errors that were not taken into account during the assembling and calibration of the Pantheon sample. At the same time, it could also hint at a presence of a physical process that causes such a redshift correlated deviation from the $\Lambda$CDM model cosmological distance evolution.
We would like to 
re-emphasize that, in this work, we are not aiming to distinguish these two possible solutions, but rather test them simultaneously. 

Unlike the Pantheon sample, residuals from the Pantheon+ sample baseline analysis shown in the right panel of \Cref{fig:res} do not clearly feature this kind of correlation.
This appears in agreement with lower correlation present in the covariance matrix shown on the right-hand panel of \Cref{fig:covmats}.
It might be caused by the change in the approach used for the estimation of the covariance in the Pantheon+ sample. By incorporating individual supernova (SNe) lightcurves, the Pantheon+ sample achieves a more accurate estimate of the covariance, in contrast to the overestimated covariance resulting from the use of binned data in the original Pantheon sample \cite{2021ApJ...912L..26B}. 
The Pantheon collaboration reports a lower significance of cosmological parameter redshift evolution \cite{ref:pan2022}. For completeness, we perform a search for additional covariance in the Pantheon+ sample data as well.

\begin{figure}
     \centering
     \begin{subfigure}{0.495\textwidth}
         \centering
         \includegraphics[width=\textwidth]{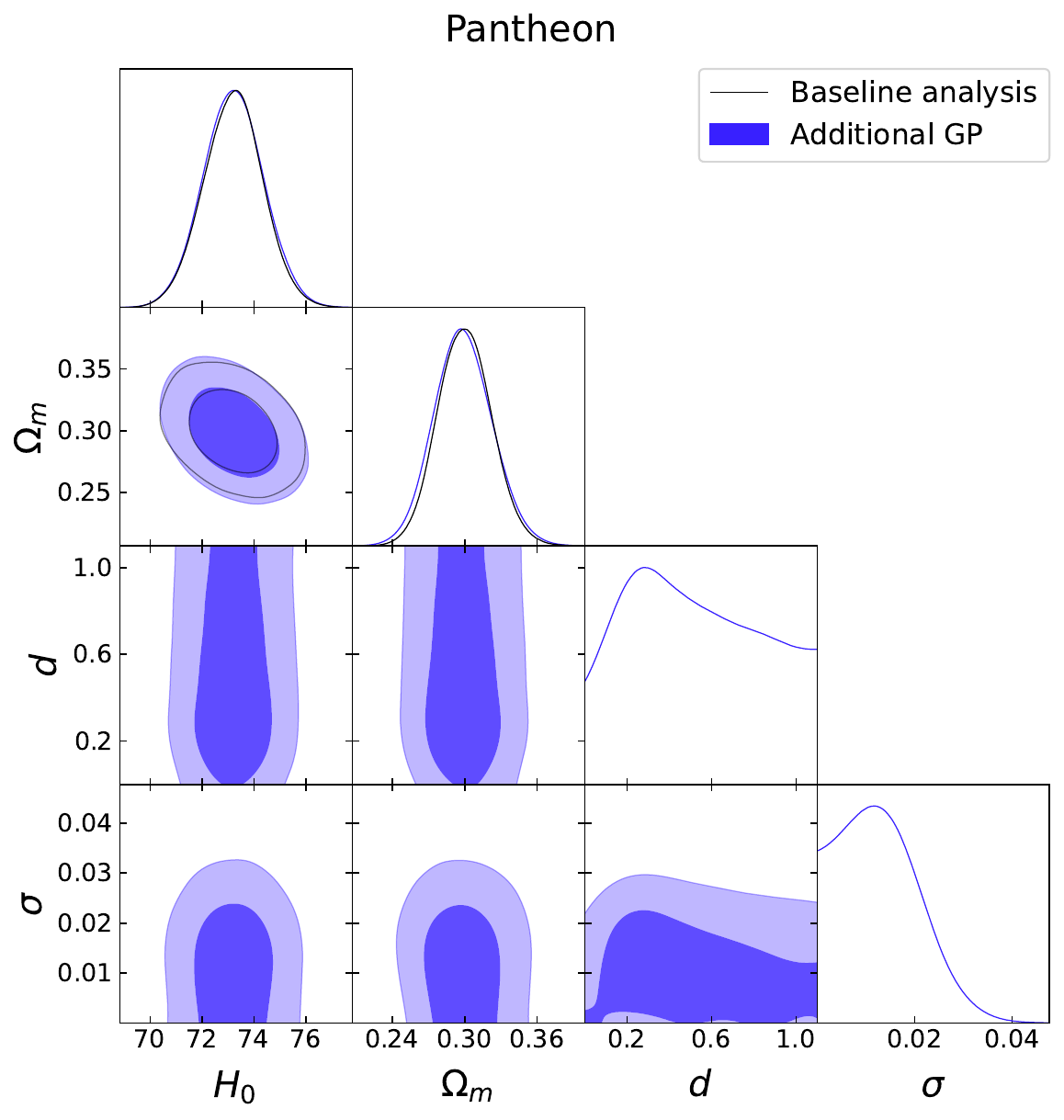}
     \end{subfigure}
     \hfill
     \begin{subfigure}{0.495\textwidth}
         \centering
         \includegraphics[width=\textwidth]{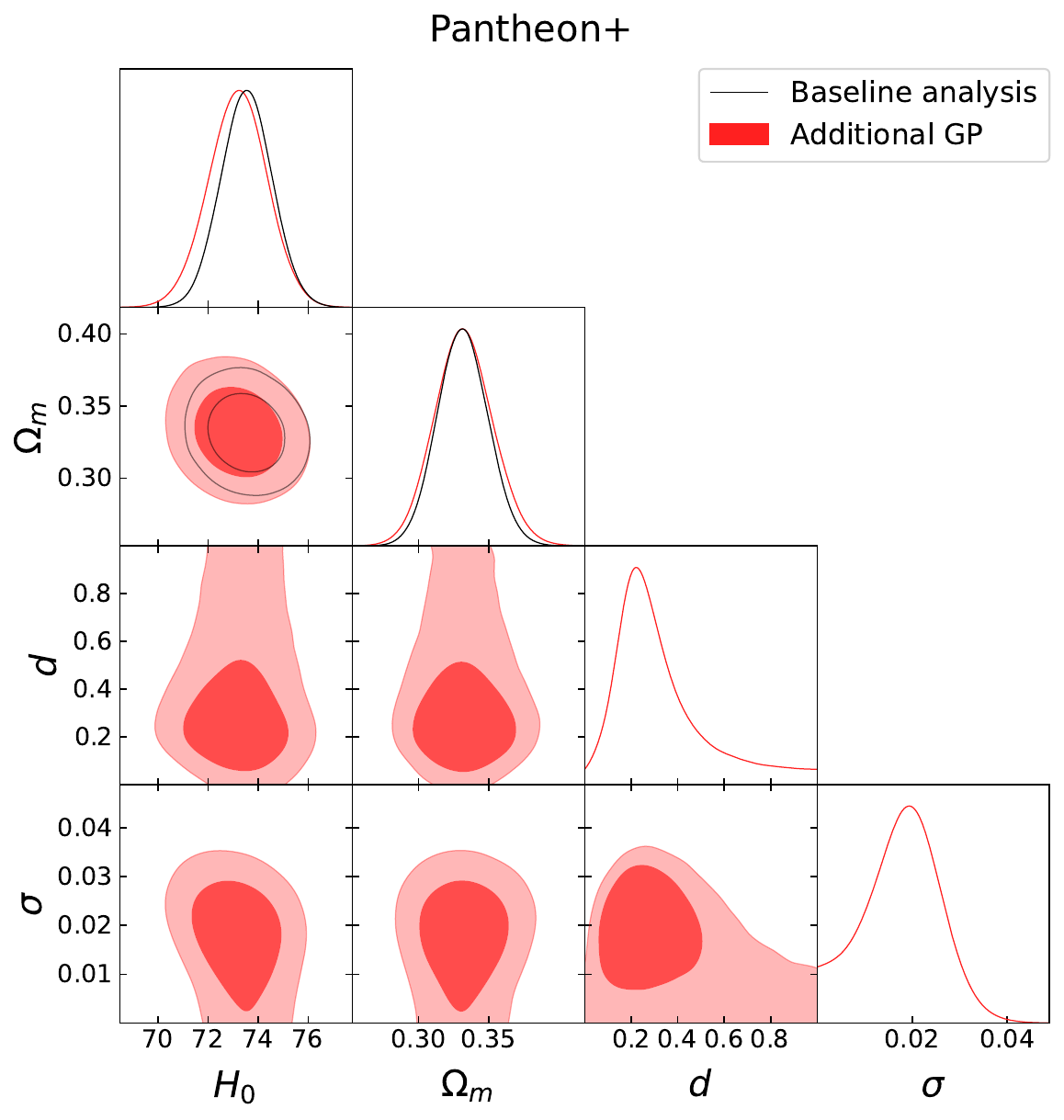}
     \end{subfigure}
        \caption{ \label{fig:post:gpvsnogp} Constraints on the flat $\Lambda$CDM cosmological model parameters $\Omega_m$, $H_0$ and parameters of the additional Gaussian process characteristic scale parameter $d$ and the amplitude parameter $\sigma$. Left and right panels represent constraints from the Pantheon and Pantheon+ samples, correspondingly. Black contours represent constraints obtained with baseline analysis that does not include the additional GP.}
\end{figure}

\subsection{Additional covariance constraints}
Constraints on the additional covariance model are presented as 68\% and 95\% confidence regions of marginalized two-dimensional posterior distributions in \Cref{fig:post:gpvsnogp}. On the left panel we present constraints obtained using the Pantheon sample and on the right panel we show the constraints from the Panthon+ sample. The black solid line contours represent constraints on cosmological parameters obtained with the baseline analysis that are in full agreement with the original Pantheon and Pantheon+ constraints. There is little impact of the additional GP on the posterior distribution of the cosmological parameters. Parameter constraints for all models are summarized in \Cref{tab:2}.
\begin{table}[]
\centering
\begin{tabular}{ |c|c|c|c|c|c| } 
 \hline
 Name & $\Omega_m$ & $H_0,\mathrm{\ km}/\mathrm{s}/ \mathrm{Mpc}$ & $\sigma$ & $d$ & $\log\mathcal{Z}$ \\ 
 \hline
 Pantheon &  $0.300\pm 0.022$  & $73.20\pm 1.13$ & \textemdash &\textemdash & $1452.1 \pm 0.3$\\ 
 Pantheon GP & $0.299\pm 0.024$ & $73.21\pm 1.16$ &  $0.0130^{+0.0051}_{-0.011} $ & $0.53^{+0.25}_{-0.42}$& $1451.1 \pm 0.3$ \\ 
 Pantheon+ & $0.332\pm 0.018$ & $73.55\pm 1.02$ & \textemdash & \textemdash & $2410.8 \pm 0.3$\\ 
 Pantheon+ GP & $0.333\pm 0.021$ & $73.21\pm 1.17$ & $0.0178^{+0.0091}_{-0.0075}$ & $0.41^{+0.11}_{-0.33}$& $2410.4 \pm 0.2$ \\ 
 Pantheon+ b.f. GP & $0.333\pm 0.02$ & $73.23\pm 1.14$ & $0.0182$ & $0.184$& $2410.6\pm 0.3$ \\ 
 \hline
\end{tabular}
\caption{\label{tab:2} Mean and 68\% confidence intervals of inferred model parameters and logarithmic Bayesian evidence.}
\end{table}

\subsubsection{Pantheon}
In the Pantheon sample case, the amplitude of the additional GP parameter $\sigma$ is consistent with zero and shows minimal correlation with the baseline model parameters. It is in agreement with a negative change in Bayesian evidence $\Delta \log\mathcal{Z} = -0.94\pm 0.45$ , which means that it is not preferred over the baseline model.
This means that the residuals, as well as the apparent redshift evolution of the parameter estimates, do not require additional covariance.
The data is consistent with the baseline model and its number of degrees of freedom. Currently, it does not support the idea of redshift dependence of cosmological parameters.

Values of the cosmological parameters inferred including the additional GP are consistent within the sub-percent region of the baseline parameter inference uncertainties and the posterior uncertainties are affected by less than $10\%$.
The most interesting feature of the obtained posterior is the presence of a small non-zero peak in the posterior distribution of the characteristic scale $d$ , with a peak at $d\approx0.18$. Due to the low evidence for the presence of any GP in the data, however, this scale is poorly constrained. The $\nu$ parameter, which sets the shape of the Mat\'ern kernel, is not constrained within our prior range. It does not have any effect on the cosmological parameters. This could perhaps suggest applying a simpler GP kernel, however, as we will show in \Cref{app:altk}, the use of simpler kernels does not improve constraints nor increases the Bayesian evidence.

\subsubsection{Pantheon+}
Analysis of the Pantheon+ sample data with an additional GP shows no statistically significant change in the Bayesian evidence, $\Delta \log\mathcal{Z} = -0.43\pm 0.41$, which, as for Pantheon, implies no preference between baseline analysis and the additional covariance model.
Despite a less prominent correlation feature in the residuals of the Pantheon+ data, the characteristic length scale $d\approx0.18$ still appears in the posterior distribution of $d$. This scale is better constrained due to higher evidence ratio $\Delta \log\mathcal{Z}$, but still not very significant. 
We highlight the part of the covariance matrices that correspond to SNe that are separated by $\Delta z<0.18$ with yellow color in \Cref{fig:covmats}. This scale coincides with the scale of the high correlation block structure around the diagonal of the covariance matrices.
Similar to the Pantheon case, the amplitude parameter posterior distribution is not separated from zero. The mean posterior value of $H_0$ is shifted by $<0.35\sigma$ from the originally estimated posterior value and the posterior uncertainty is $15\%$ larger.  These small changes in the posterior reduce the statistical significance of the tension between SN and CMB estimates of $H_0$ from $5.3\sigma$ to $4.5\sigma$.   

The maximum posterior value of the additional covariance amplitude corresponds to $11\pm 3\%$ of the statistical error of the observed SNe brightness and to $90\pm 90\%$ of systematic errors for the closest SN neighbors in the redshift space. 
Despite weak constraints, the Mat\'ern kernel parameters show a small degeneracy with the cosmological parameters. We tested how adding covariance with the best-fit value Mat\'ern kernel parameters affect cosmological parameter inference, and find that its effect on $\Omega_m$ and $H_0$ is similar to the effect obtained when marginalising over Mat\'ern kernel parameters. We conclude that the small shift in the Hubble parameter and the mild increase of the posterior errors presented in the last row of \Cref{tab:2} is an upper limit on impact on the Pantheon+ sample constraints of possible additional covariance.

\section{Conclusions and discussion \label{sec:concl}}
We tested the Pantheon and revised Pantheon+  SN samples for the presence of hidden Gaussian Processes in their luminosity distances that are not included in the covariance matrices released in these datasets.  
Despite a wide range of tests, we find no significant evidence for the presence of an underlying GP and, therefore, such a process cannot explain the high tension between early and late universe cosmological probes. Furthermore, the significance of the $H_0$ discordance cannot be lowered below $4.5\sigma$ significance from the 5.3 $\sigma$ for the baseline model within the obtained constraints on the GP parameters. 
In \Cref{app:nonstat} and \Cref{app:altk}, we tested our assumptions on the stationarity of GP and the selection of the particular type of GP kernel and found that lifting those assumptions does not change our conclusion either.   

While visually stronger hints for the redshift evolution of cosmological parameters are present in the Pantheon sample, contrary to this, we found higher evidence for the presence of additional GP in the Pantheon+ sample. One of the possible explanations is the inclusion of a full covariance matrix in our analysis, contrary to the redshift binning techniques used by Ref. \cite{ref:zev}, which do not take into account parts of the covariance describing the correlation between redshift bins.
The use of a single redshift bin also excludes part of the covariance that contains information about the anti-correlation of data in high- and low-redshift ranges \cite{ref:zmax_vs_omegam}. Therefore, we present our work as an example of an alternative approach to these kind of searches, that keep complete information present in the covariance matrix.
 
Improved GP parameter constraints and the corresponding change in the evidence ratio in the Pantheon+ dataset analysis can be a result of its higher statistical power due to a larger number of SNe in the sample, as well as a consequence of improvements in the computation of the systematic errors.     
The higher impact of the additional GP on cosmological parameter inference with the Pantheon+ sample can be explained by the inclusion of calibration SNe to the dataset with the complete covariance. Thus, GP parameters are more correlated with cosmological parameters in this case. The narrow localisation of calibration SNe in redshift space makes SN probes more sensitive to model extensions with additional redshift-dependent GPs. 

Despite the lack of substantial evidence for additional covariance, the presence of a characteristic scale can motivate further studies of the SN sample covariance. The coincidence of the block structure of the Pantheon+ covariance with the characteristic scale highlighted in \Cref{fig:covmats} can serve as motivation for more detailed studies of the systematic error budget of the baseline Pantheon+ analysis.

Further studies of this sort can test the impact of hidden GP processes that are correlated in different parameters of the SNe, such as the color and stretch of the lightcurve, the mass of host galaxy, etc. 
For example, the position of SN in the sky is one of the interesting parameters to study in the context of recent discussions on the difference between local and CMB measured kinematic dipole  \cite{ref:dipolepanp,ref:dipoleradio}. Upcoming studies can also include less processed data and consider additional GP during the systematic-error estimation phase.

\acknowledgments
B.B. is supported by the Fundamentals of the Universe research program within the University of Groningen. P.D.M. acknowledges support from the Netherlands organization for scientific research (NWO) VIDI grant (dossier 639.042.730). 

Contours and parameter constraints are generated using GetDist \cite{ref:getdist}. Plots are generated with  Matplotlib \cite{ref:matplotlib}. We use SciPy \cite{ref:scipy}, Astropy \cite{ref:astropy}, NumPy \cite{ref:numpy}, pandas \cite{ref:pandas}, and scikit-learn \cite{ref:scikit-learn}.
\appendix
\section{\label{app:nonstat} Nonstationary Gaussian Process }
In the main part of our analysis, we test the Pantheon data for the presence of additional covariance assuming that it can be described by a stationary Gaussian kernel, i.e. independent of redshift. Here, we test the robustness of the obtained results by introducing non-stationarity in two ways: 1) applying a smooth redshift-dependent scaling for the amplitude of the additional covariance $\sigma(z)$ and 2) considering only a fraction of the data and covariance matrices with an upper redshift cut.
\begin{figure}
     \centering
         \includegraphics[width=\textwidth]{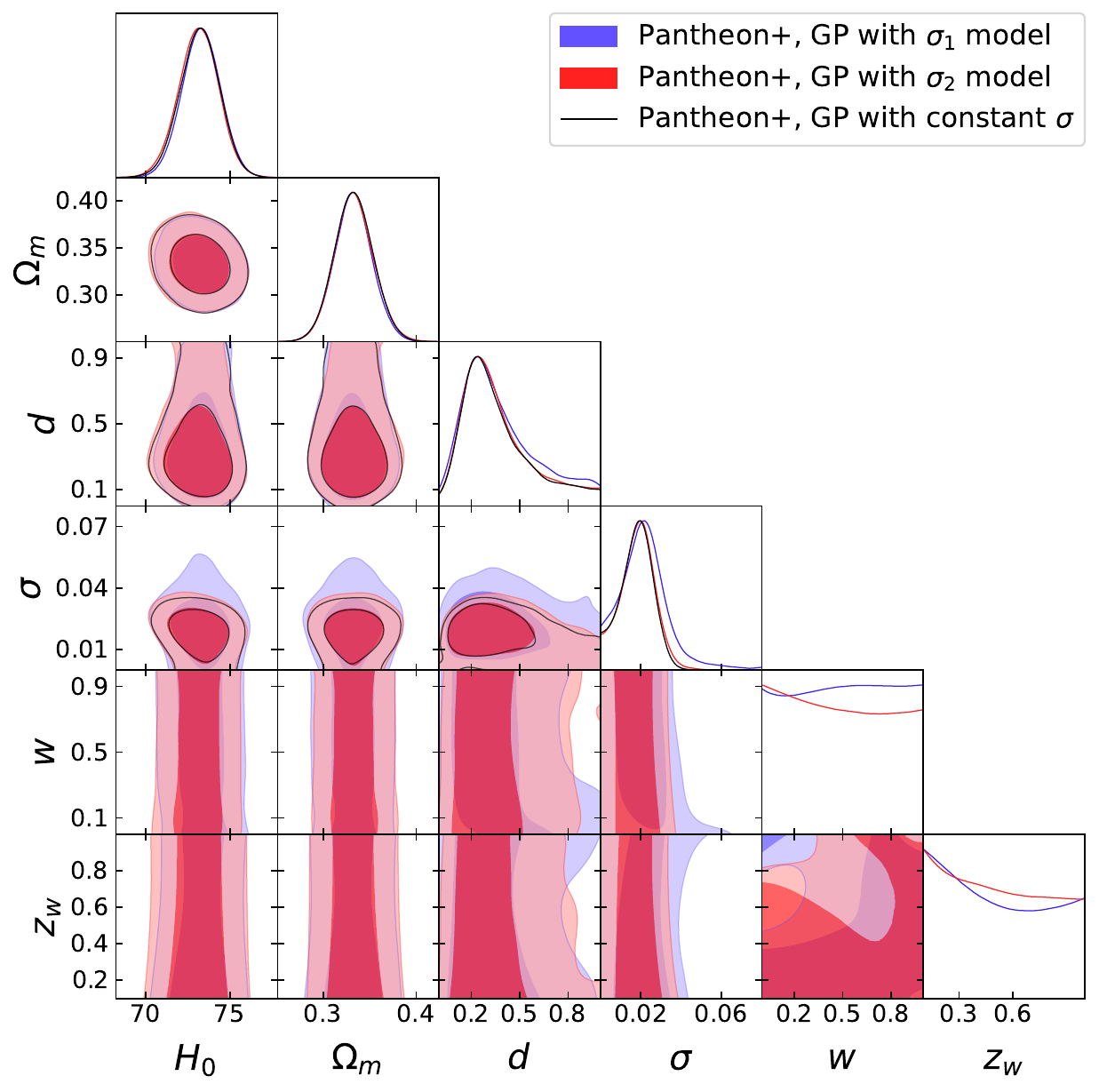}
        \caption{Parameter constraints using GP models that include redshift evolution of the GP amplitude. The blue contour represents the model described in \Cref{eq:s1}, the red contour represents the model described in \Cref{eq:s2}, and the black contour represents the GP model used in the main text.}
        \label{fig:a1sigmas}
\end{figure}
We test these two realizations of the amplitude redshift dependencies, replacing the redshift-independent $\sigma^2$ factor in \Cref{eq:MatKer}.
The first one is a Gaussian function:
\begin{equation}  \label{eq:s1}
    \sigma_1^2(z,\sigma,w,z_w) =\sigma^2 \exp\left(-\frac{(z-z_w)^2}{w^2}\right),
\end{equation}
where parameter $z_w$ defines the redshift for which the additional GP has the strongest effect, and $w$ defines the rate of the GP amplitude decrease away from $z_w$. The inferred parameter constraints on this model extensions are shown in blue in \Cref{fig:a1sigmas}. 
The second realization has constant amplitude of the GP at low redshifts which decreases as a Gaussian function after a certain redshift limit $z_w$: 
\begin{equation} \label{eq:s2}
    \sigma_2^2(z,\sigma,w,z_w) =\begin{cases} \sigma^2, & \mbox{for } z < z_w \\
\sigma^2 \exp\left(-\frac{(z-z_w)^2}{w^2}\right), & \mbox{for } z\geq z_w .
\end{cases}
\end{equation}
Using this GP, we find no evidence for any preferred realization of redshift evolution in the GP. The additional parameters are not constrained beyond the priors and do not affect cosmological parameter inference. 

\begin{figure}
     \centering
     \begin{subfigure}{0.325\textwidth}
         \centering
         \includegraphics[height=\textwidth]{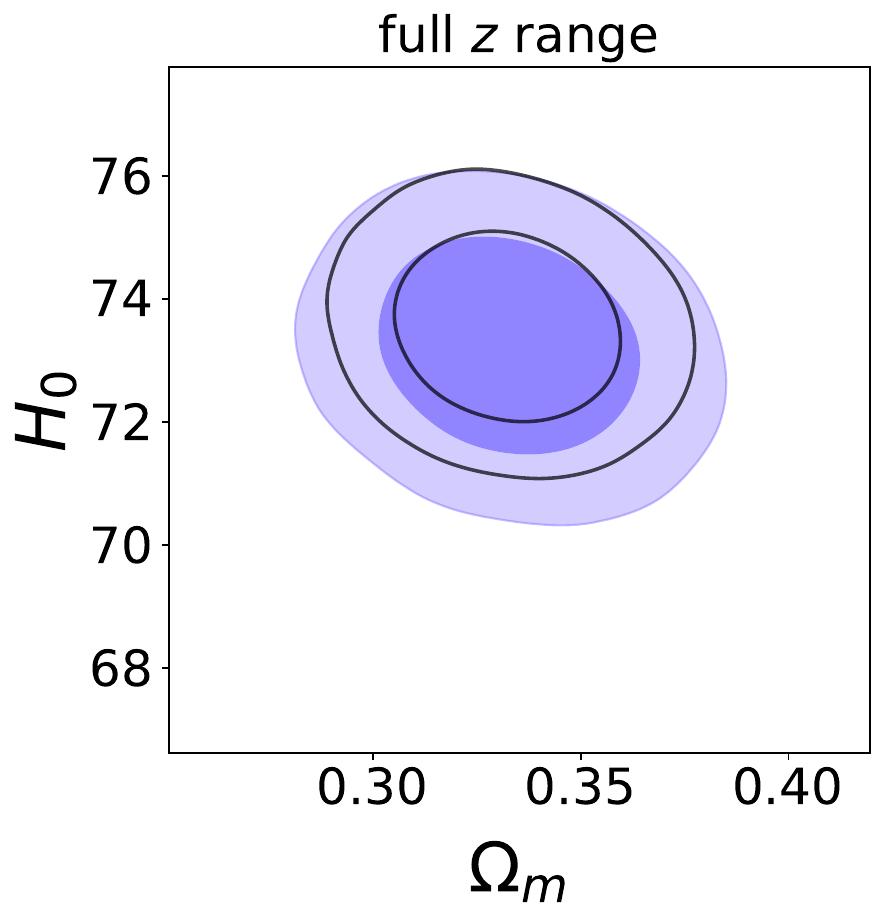}
    \end{subfigure}
     \hfill
    \begin{subfigure}{0.325\textwidth}
         \centering
         \includegraphics[height=\textwidth]{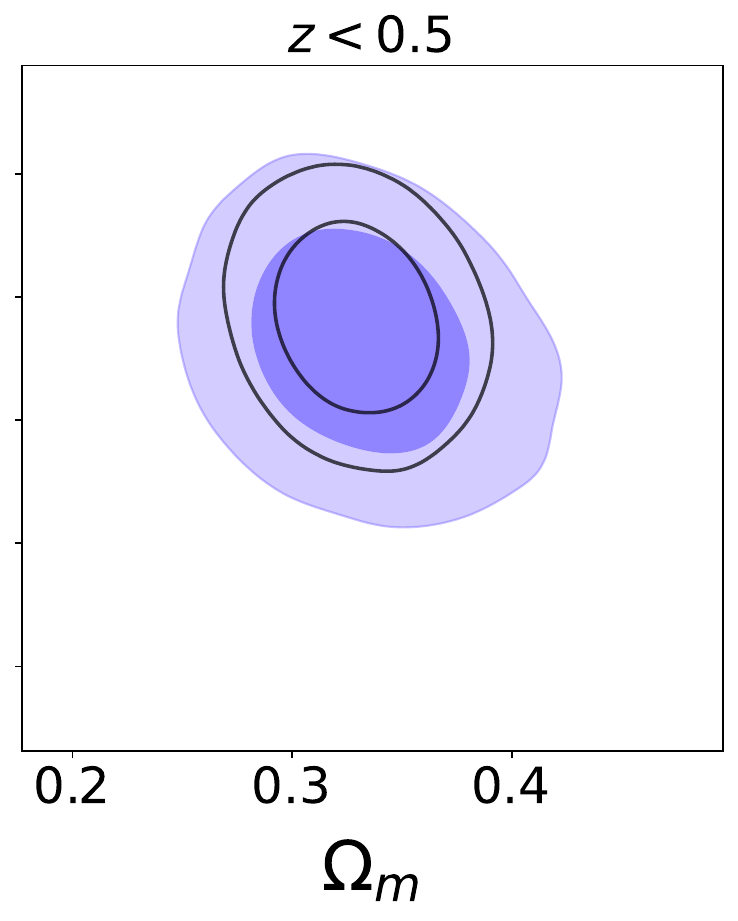}
    \end{subfigure}
    \begin{subfigure}{0.325\textwidth}
         \centering
         \includegraphics[height=\textwidth]{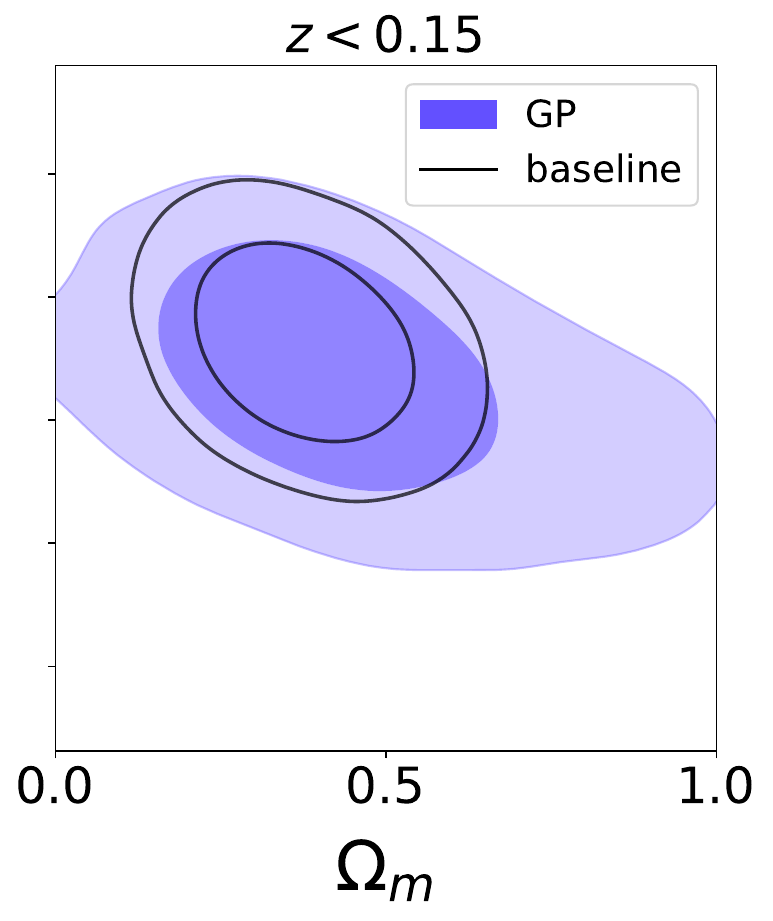}    
    \end{subfigure}
        \caption{Comparison of cosmological parameter constraints obtained from different redshift cuts of the Pantheon+ sample with (blue filled contours) and without additional Gaussian process (black contours). All panels share a vertical axis scale representing Hubble parameter constraints while the horizontal axis scale is different for each panel. The left panel shows constraints obtained with the full redshift range of SNe, the central panel constraints are obtained with SNe within $z<0.5$ range, and the right one is obtained with SNe in $z<0.15$ range.}
        \label{fig:a12}
\end{figure}

Next, we consider two redshift cuts on the upper limit of the used SNe redshifts. The first limit is $z<0.15$, which is similar to the redshift upper limit used by the SH0ES collaboration. Information on SN distances below that redshift contains most of the constraining power on the Hubble parameter \cite{ref:sh0es2016,ref:sh0es2022,ref:pan2022}. Constraints are shown in the right panel in \Cref{fig:a12}.
The second limit that we explore is $z<0.5$. We chose this redshift range due to stronger constraints on $\Omega_m$ and a reported weak redshift dependence \cite{ref:pan2022}. These constraints are shown in the middle panel in \Cref{fig:a12}.
For both redshift bins, we find no improvements in constraints of the GP parameters and no change in impact on $H_0$ inference. 
A small increase in the effect on $\Omega_m$ inferred posterior is expected due to the weaker constraints of this parameter within the low-redshift window and stronger sensitivity to data uncertainties.

We conclude that these non-stationary extensions of the GP model do not change our conclusion and do not change cosmological parameter constraints. Tests with the original Pantheon sample show similar results, with no evidence for non-stationarity.   
    
\section{\label{app:altk} Alternative kernel}
\begin{figure}[h]
     \centering
         \includegraphics[width=\textwidth]{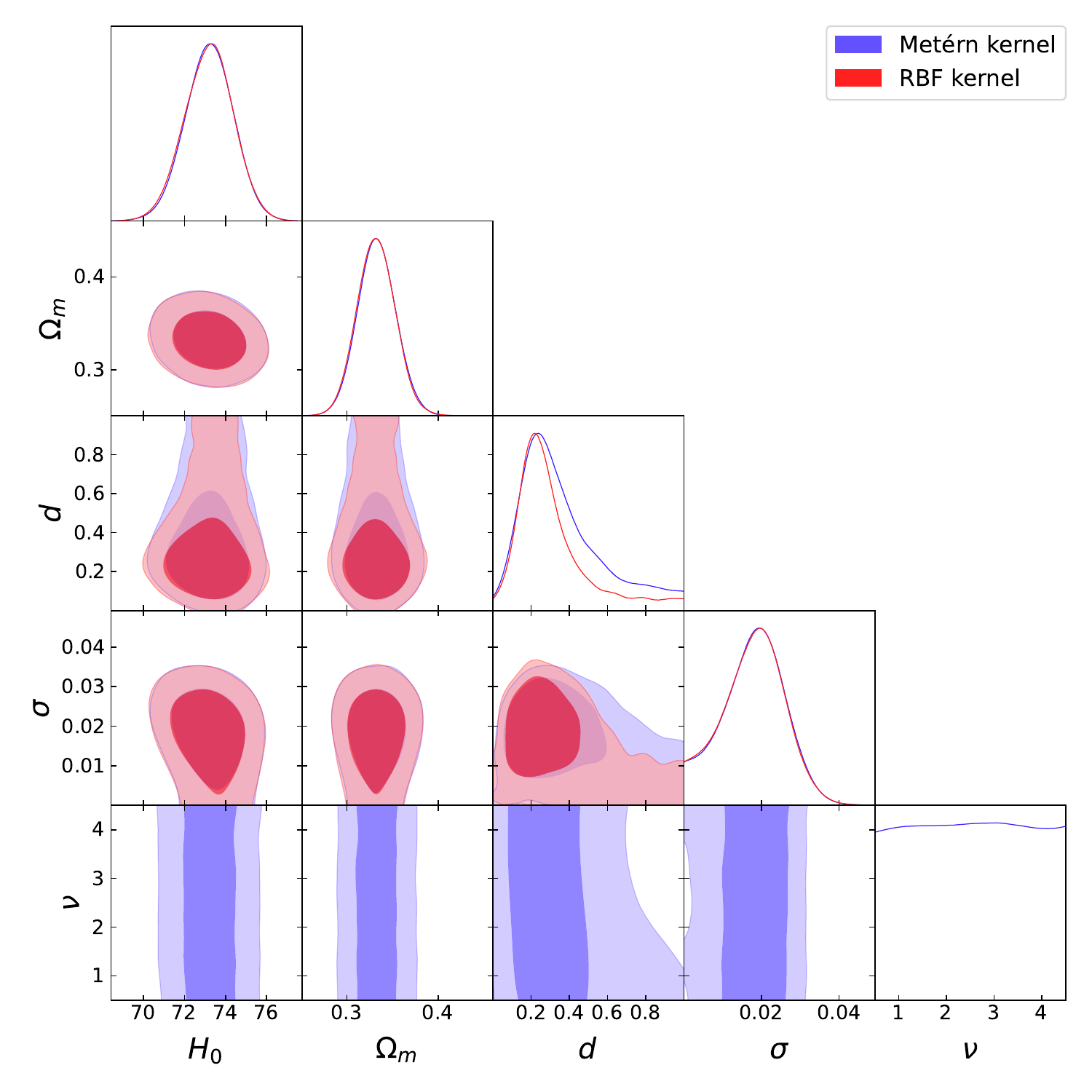}
        \caption{Comparison of parameter constraints obtained with Pantheon+ data for two model of additional covariance: Mat\'ern kernel (blue) and  RBF kernel (red).}
        \label{fig:RBF}
\end{figure}

We use the Mat\'ern kernel as our main kernel for the additional covariance for the flexibility and variety of possible covariance shape realizations. 
However, it could be sub-optimal, and other kernels might perform better. 
Moreover, simpler kernels might be preferred due to a smaller number of additional parameters. 
Here we present constraints on a simpler covariance model - the Radial basis function kernel:
\begin{equation}
    \label{eq:RBF}
    C_{\mathrm{RBF}}(\Delta z) = \sigma^{2} \exp{\left(- \frac{{\Delta z}^2}{2d^2}\right)}.
\end{equation}
This kernel is equivalent to the Mat\'ern kernel when $\nu \to \infty$. Despite having one parameter less, it does not result in improved Bayesian evidence with an estimated value of $\log \mathcal{Z} = 2409.7\pm 0.3$, a difference of $\Delta \log \mathcal{Z}  = -1.1$ with respect to the baseline analysis. A comparison of parameter constraints obtained for Pantheon+ data with the Mat\'ern and the RBF kernels is shown in \Cref{fig:RBF}. The RBF kernel has a similar effect on the cosmological parameter inference as the Mat\'ern kernel.
\newline

\bibliographystyle{JHEP}

\bibliography{ref.bib}

\end{document}